\documentclass{ipart_v1}

\usepackage{t1enc}
\usepackage[latin1]{inputenc}
\usepackage[english]{babel}

\usepackage{amsthm}
\usepackage{yfonts}

\usepackage{bbm}
\usepackage{bm}
\usepackage{mathrsfs}
\usepackage{graphicx}
\usepackage{ragged2e}

\newcommand{\be}[0]{\begin{equation}}
\newcommand{\ee}[0]{\end{equation}}
\renewcommand{\raggedright}{\leftskip=0pt \rightskip=0pt}

\numberwithin{equation}{section}

\theoremstyle{plain}

\begin{document}

\title[Paradox in  relativity]{Four paradoxes about the special theory of relativity}

\author[Xiao-Jun Li et al.]{Xiao-Jun Li$^{1,*}$, Kai Chen$^1$, Xiao-Niu Li$^2$, Yong-An Li$^3$
 \address{\textsl{\small \\$^1$Institute of Photonics \& Photon-Technology, Northwest University, Xi'an 710069, China; \\$^2$Shaanxi Radio \& TV University, Xi'an 710119, China; \\$^3$Shaanxi BC \& TV Network Intermediary Co. Ltd. Yan'an Branch, Yan'an 716000, China; \\$^*$Corresponding author, \email{lixiaojun@nwu.edu.cn}}}}

\begin{abstract}
Various paradoxes about the relativity theory have been developed since the birth of this theory. Each paradox somewhat shows people's query about the relativity theory, and solving of each paradox demonstrates the correctness of relativity theory once again. In this paper, four paradoxes about the special theory of relativity are brought forward: displacement paradox, electromagnetic transformation paradox, Doppler paradox and magnetic force paradox. We hope some researchers can reasonably explain these paradoxes, and then knowledge of the relativity theory will become more abundant.
\end{abstract}

\maketitle

\section{Introduction}

The theory of relativity has been established for more than 100 years \cite{Einstein: 1905}. The theory shattered the framework of classical physics \cite{Hogg: 1997, Einstein: 1916} with its unprecedented results such as time dilation, length contraction, non-conservation of mass, the universal speed limit of light, space curvature, etc. Until now, relativity has been widely accepted by the physics community, and some of its predictions have been experimentally confirmed to some degree \cite{Zhang: 1997, Hobson: 2005}.

However, a variety of paradoxes were derived from the relativity theory, such as twin paradox \cite{Grunbaum: 1954}, slide block paradox \cite{Rindler: 1961}, submarine paradox \cite{Matsas: 2003}, right-angled lever paradox \cite{Tolman: 1934}, soft rope paradox \cite{Bekesy: 1955} and so on. These paradoxes show people's deep concern about the theoretical system of the relativity, at the same time, this also reflects some misgivings about the relativity. On the other hand, as the paradoxes were resolved one by one, the theory system of the relativity became more abundant, more perfect, and more convincing.

In this paper, four paradoxes about the special theory of relativity are put forward. (i) Displacement paradox: the displacement of points depends on the arbitrary localized coordinate origin during the length contraction. (ii) Electromagnetic transformation paradox: the transformation of electromagnetic field seems to contradict with the relativistic principle. (iii) Doppler paradox: the relativistic Doppler effect seems to conflict with the chasing light thought experiment. (iv) Magnetic force paradox: there is no magnetic interaction between moving electrons. We hope that putting forward and solving of these paradoxes can contribute to the theory of relativity.

\section{Displacement paradox}

\label{SecRemDiffCoc}

According to the Lorentz transformation:
\be
x'=\frac{x-vt}{\sqrt{1-v^2/c^2}}
\ee
This equation leads to
\be
x=x'\sqrt{1-v^2/c^2}+vt
\ee
For simplicity, only the case when $t=0$ is considered here, then,
\be
x=x'\sqrt{1-v^2/c^2}
\ee
At this moment, the origin of the static reference frame K and moving reference frame K$'$ coincide with each other, as shown in Fig. 1.

\begin{figure}[h]
\centerline{\includegraphics[width=57mm]{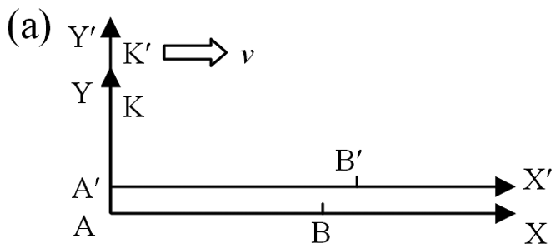}~~~
 \includegraphics[width=57mm]{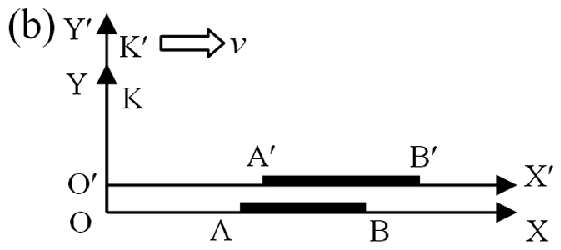}}
\begin{flushleft}
    \textbf{Figure 1.} Displacement in the length contraction. (a) Point A$'$ is at the origin, so point A coincides with A$'$. Point B$'$ is  not at the origin, so point B has a displacement toward the origin. (b) Rod A$'$B$'$ is separated from the origin, and rod AB has a displacement toward the origin relative to rod A$'$B$'$.
\end{flushleft}
\end{figure}

In frame K$'$, a point A$'$ is placed at the origin of coordinate, $x'_A=0$. According to (2.3), $x_A=0$, so point A (the corresponding point of A$'$) in frame K is also at the origin, i.e., point A coincides with A$'$. See Fig. 1(a).

Another point B$'$ on the X-axis of frame K$'$ is not at the coordinate origin; for instance, $x'_B>0$, so the position of point B in frame K is $x_B=x'_B\sqrt{1-v^2/c^2}<x'_B$ according to (2.3), i.e., B does not coincide with B$'$. B has a displacement toward the origin relative to B$'$. The displacement is proportional to the value of $x'_B$. See also Fig. 1(a).

It can be conclude here that, every point in frame K will have a displacement toward the origin relative to its corresponding point in frame K$'$, and the displacement depends on |x$'$|. If x$'$=0, there is no displacement. The larger |x$'$| is, the further the point is displaced. It is known that a certain point's coordinate value x$'$ depends on the localization of the origin. However, the position of the origin is determined arbitrarily. When the origin is placed at different positions, the points in frame K will be displaced toward different positions. For example, if the origin is localized at A$'$, A remains fixed and B is displaced toward A; if the origin is localized between A$'$ and B$'$, A and B will be displaced toward a position between they two. This phenomenon that the displacement depends on the arbitrarily localized coordinate origin seems to be unreasonable.

For a rod A$'$B$'$ with a certain length in frame K$'$, as shown in Fig. 1(b), not only will its length contract but its entire body will also be displaced toward the origin. If rod A$'$B$'$ is placed on the positive X-axis, AB will be displaced leftward relative to A$'$B$'$; if rod A$'$B$'$ is on the negative X-axis, AB will be displaced rightward. The further A$'$B$'$ is from the origin, the greater the displacement is. In the extreme case of $v=c$, for any rod along the X-axis in frame K$'$, not only will its length contract to zero, but its entire body also will be compressed into the origin while being observed in frame K. However, the origin of coordinate is arbitrarily localized. This phenomenon seems to be unreasonable.

Let's assume a metal bar be in equilibrium state at zero absolute temperature. When the bar is heated to a very high temperature, the free electrons inside the metal will undergo drastic thermal motion (the thermal motion of the metal lattice can be neglected). Due to the high-speed thermal motion, the free electrons will experience space contraction. Not only will the size of the electrons contract, but also will the distance between the electrons diminish. In fact, according to (2.3), the free electrons will have a certain displacement as a whole toward the origin of coordinate. As shown in Fig. 2, if the metal bar is placed on the right of the origin, the free electrons will be displaced leftward, then, the bar will have positive charges on its right end and negative charges on its left end (the length contractions in Y and Z directions are neglected here). In the opposite case, if the metal bar is placed on the left of the origin, the free electrons will be displaced rightward, then, the bar will have positive charges on its left end and negative charges on its right end. If the metal bar is placed just at the origin, the free electrons will contract inward without displacement as a whole, and the bar will have positive charges on both ends and negative charges at the center. Is it reasonable that the electrical property of the metal bar is dependent on the arbitrarily localized coordinate origin?

\begin{figure}[h]
\centerline{\includegraphics[width=85mm]{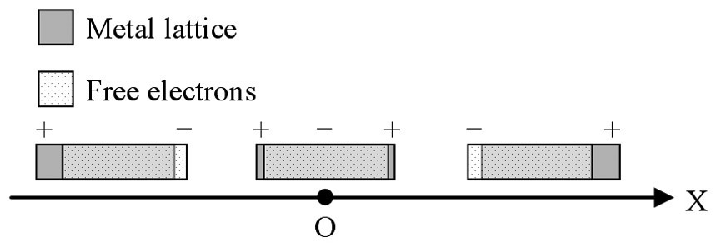}}
\begin{flushleft}
    \textbf{Figure 2.} Space contraction of free electrons in a metal bar. High-speed thermal motion causes length contraction as well as displacement of the free electrons as a whole. If the metal bar is on different side of the origin, the free electrons will have displacement in different directions. This will lead to different electrical properties of the metal bar.
\end{flushleft}
\end{figure}

Furthermore, if there is no coordinate system, we will completely no way to tell which direction the free electrons will be displaced toward as a whole, how much the displacement is, or they just contract inward. Is this reasonable?

The concomitant displacement in the length contraction is seemingly unreasonable. We call this phenomenon as displacement paradox.

\section{Electromagnetic transformation paradox}
\label{SecGlobAn}

There is a uniform magnetic field {\bf B} in the upward direction. From the view of a static observer K, there is only a static magnetic field with $B_y>0$, $B_x=B_z=0$, {\bf E}=0, as shown on the left in Fig. 3.

\begin{figure}[h]
\centerline{\includegraphics[width=85mm]{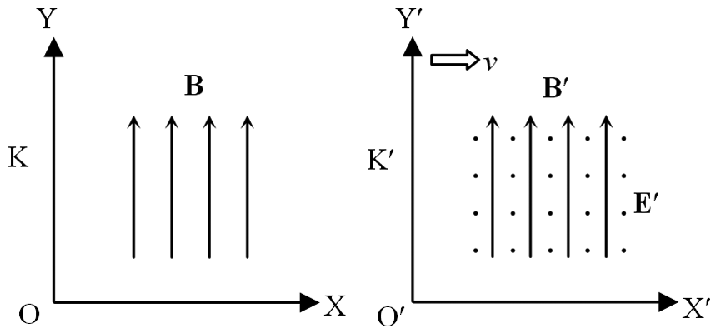}}
\begin{flushleft}
    \textbf{Figure 3.} Electromagnetic field from the view of different observers. Magnetic field {\bf B} in the eyes of static observer K becomes {\bf B$'$} and {\bf E$'$} from the view of moving observer K$'$. K only feels magnetic force while K$'$ feels electrical force and magnetic force.
\end{flushleft}
\end{figure}

Another observer K$'$ moves at a velocity $v$ relative to K (along the positive X-axis direction). From the view of K$'$, there is not only magnetic field but also electric field (see the right of Fig. 3). {\bf E$'$} and {\bf B$'$} can be deduced according to electromagnetic field transformation formula \cite{Griffiths: 1999} in special relativity as follows:
\be
E'_z=\frac{E_z+vB_y}{\sqrt{1-v^2/c^2}}=\frac{vB_y}{\sqrt{1-v^2/c^2}}
\ee
\be
B'_y=\frac{B_y+v/c^2 E_z}{\sqrt{1-v^2/c^2}}=\frac{B_y}{\sqrt{1-v^2/c^2}}
\ee
The other components of {\bf E$'$} and {\bf B$'$} are all zero. It can be seen here that, based on the theory of relativity, observers in different moving states may see different electromagnetic fields.

Let the two observers K and K$'$ keep their moving states unchanged, forbid information exchange between them, and confiscate the former calculation result of K$'$. Under this circumstance, if K$'$ wants to know the electromagnetic field value, he will not be able to calculate using (3.1) and (3.2), and the only way he can rely on is measuring.

Now, let's carry out a measurement in mind. A small magnetic needle is held in the left hand of observer K and a small charged ball in his right hand. Then, his left hand will feel a magnetic force, while the right hand will not feel any force (the charged ball does not move relative to K, so there is no Lorentz force). K could obtain the value of the magnetic field {\bf B} by observing the deflection of the needle in his left hand. Meanwhile, same magnetic needle and charged ball are held in the left and right hands of observer K$'$  respectively. Then, his left hand will feel a magnetic force that is larger than what K feels (because  $B'_y>B_y$) and his right hand will feel an electric force simultaneously. Similarly, K$'$ could obtain the value of the electric field {\bf E$'$} and magnetic field {\bf B$'$} according to the forces he feels.

However, a serious problem arises here. K and K$'$ are completely equivalent observers. They are at equal status and under equal circumstance. K$'$ moves relative to K, and K also moves relative to K$'$ in the same way. There is no any difference between them. How can the two completely equivalent observers obtain different values of the electromagnetic field? Why do their corresponding hands feel entirely different electromagnetic forces? There is no other thing around K and K$'$ except the field. You are no way to show that K and K$'$ are at different states, unless you can prove that their movement modes are different relative to the field. For example, K is static relative to the field while K$'$ moves relative to the field, or on the contrary, K$'$ is static while K moves relative to the field. However, concept of ``being static'' or ``moving'' relative to the field is not permitted in the relativity theory. The field can only ``exist'', but cannot be ``static'' or ``moving''. It has no concept of moving state and cannot be used as reference frame. Thus, there is no reason to say that the states of K and K$'$ are not equal.

The problem can also be described and analyzed as below.

As shown in Fig. 4, there is an electromagnetic field ({\bf E}, {\bf B}). A charge Q$_1$ is fixed in the static reference frame K. Then Q$_1$ will experience an electromagnetic force {\bf F}$_1$. From view of the moving frame K$'$ (with a velocity {\bf v}), the electromagnetic field becomes ({\bf E$'$}, {\bf B$'$}). Considering that Q$_1$ moves at a velocity -{\bf v} relative to frame K$'$, in order to keep the electromagnetic action Q$_1$ experiences unchanged in K$'$, generally there should be {\bf E$'$}$\not=${\bf E} and {\bf B$'$}$\not=${\bf B}.

\begin{figure}[h]
\centerline{\includegraphics[width=85mm]{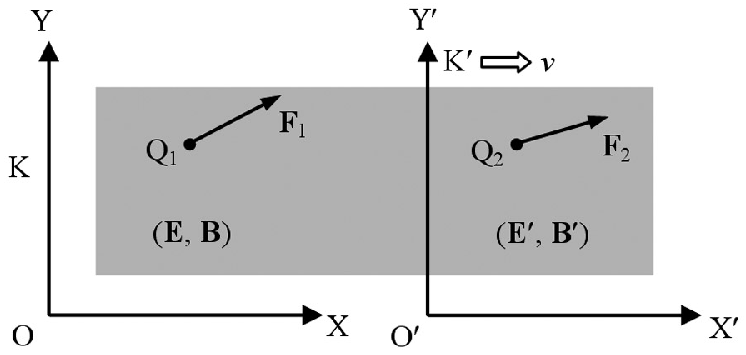}}
\begin{flushleft}
    \textbf{Figure 4.} Reference frame transformation of electromagnetic field and the relativistic principle. The electromagnetic field ({\bf E}, {\bf B}) in frame K becomes a different electromagnetic field ({\bf E$'$}, {\bf B$'$}) in frame K$'$. The identical charge Q$_1$ and Q$_2$ which are fixed in frame K and K$'$ respectively will experience different forces.
\end{flushleft}
\end{figure}

Meanwhile, a charge Q$_2$ which is completely identical to Q$_1$ is fixed in frame K$'$. It will experience an electromagnetic force {\bf F}$_2$. Because {\bf E$'$}$\not=${\bf E} and $Q_2=Q_1$, there will be {\bf F}$_2\not=${\bf F}$_1$ (because Q$_1$ and Q$_2$ are static relative to their own frames, they do not experience Lorentz force).

A problem arises here. The two identical charges Q$_1$ and Q$_2$ are both static relative to their own reference frames respectively. However, they experience different electromagnetic forces. This goes against the relativistic principle. According to above analysis, just by observing the forces that Q$_1$ and Q$_2$ experience, we can find that frames K and K$'$ are at different movement states. This seems not be permitted by the relativistic principle.

In order to ensure that the physic phenomenon is independent on observer, i.e., to ensure the electromagnetic action that a single charge Q$_1$ experiences keeps unchanged in different frames K and K$'$, there should be {\bf E$'$}$\not=${\bf E} and {\bf B$'$}$\not=${\bf B} generally. This will make the two charges Q$_1$ and Q$_2$ with identical moving states in their own reference frames experience different forces. This acts against the relativistic principle. On the contrary, in order to ensure the relativistic principle not to be broken, there should be ({\bf E$'$}, {\bf B$'$})=({\bf E}, {\bf B}). This will make the Lorentz transformation disabled, and cannot ensure the physical phenomenon to be independent on observer.

According to above, the electromagnetic field transformation seems to be contradictory to the relativistic principle. We call this phenomenon as electromagnetic transformation paradox.

\section{Doppler paradox}

The Doppler effect describes the frequency change of a wave received by an observer when the observer moves relative to the wave source. The Doppler effect is a reference system effect or observational effect.

According to the special relativity theory, if there is a relative movement with velocity {\bf v} between the observer and the light source (the direction of {\bf v} is along the line connecting the observer and the light source), the light frequency measured by the observer is
\be
f_d=\sqrt{\frac{c-v}{c+v}}f_0
\ee
where $f_0$ is the initial frequency of the light. This is the relativistic Doppler effect of light \cite{Einstein: 1905}.

Now let light source keep static and the observer move along the light ray at velocity $v$. According to (4.1), if $v>0$ (the observer moves away from the light source), light frequency the observer see will decrease, i.e., $f_d<f_0$, and the larger $v$ is, the smaller $f_d$ is.

When the velocity of the observer reaches the light speed, i.e., $v=c$, then $f_d=0$. This means the light wave stops oscillating from view of the observer. In other words, the light appears as a "frozen wave" in the eye of the observer.

However, most people who are well familiar with the relativity theory may know Einstein's famous thought experiment of chasing light \cite{Einstein: 1907}. Einstein ever recalled: "If I pursue a beam of light with the velocity $c$ (velocity of light in a vacuum), I should observe such a beam of light as a spatially oscillatory electromagnetic field at rest. However, there seems to be no such thing, whether on the basis of experience or according to Maxwell's equation \cite{Schilpp: 1970}." "From the very beginning it appeared to me intuitively clear that, judged from the standpoint of such an observer, everything would have to happen according to the same laws as for an observer who, relative to the earth, was at rest. For how, otherwise, should the first observer know, i.e., be able to determine, that he is in a state of fast uniform motion? \cite{Schilpp: 1970}" Einstein here believes that, for a beam of light, all observers are equivalent, and an observer who is pursuing light will not see the light wave to be "frozen", and the Maxwell equation does not allow this either.

The thought experiment of chasing light is one of the ideological bases of the relativity theory. It can be said in a sense: without the chasing light experiment, there would be no the relativity theory.

According to the relativistic Doppler effect, when the observer runs after the light at the velocity $c$, the light frequency appears as $f_d=0$, and the light wave seems to be "frozen". However, the chasing light experiment demonstrates that the light wave will not be "frozen". Here they two seem to conflict with each other.

We call the incompatibility between the relativistic Doppler effect and the chasing light thought experiment as Doppler paradox.

\section{Magnetic force paradox}

First, consider the case of two electrons with same velocity.

As shown in Fig. 5(a), there are two electrons e$_1$ and e$_2$ that both move at the velocity $v_0$ along the positive of X-direction of the static reference frame K. Frame K$'$ moves at the same velocity $v_0$ relative to frame K. For simplicity, e$_2$ is positioned right above e$_1$, i.e., the line connecting e$_1$ and e$_2$ is parallel to the Y-axis.

Define e$_1$ as the field charge and e$_2$ as the test charge.

\begin{figure}[h]
\centerline{\includegraphics[width=57mm]{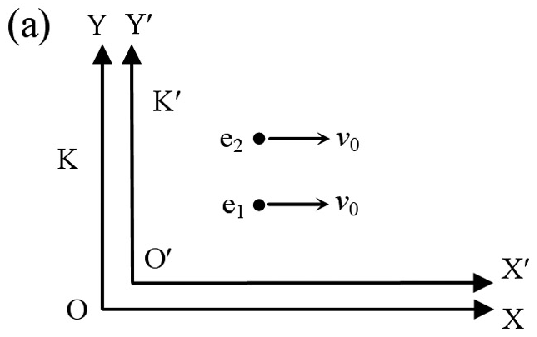}~~~
 \includegraphics[width=57mm]{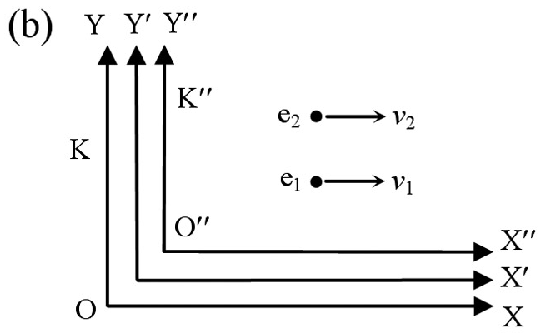}}
\begin{flushleft}
    \textbf{Figure 5.} Interaction between two electrons. (a) The two electrons move at the same velocity $v_0$. K is the static frame, and K$'$ is the moving frame with velocity $v_0$. (b) The two electrons move at different velocities $v_1$ and $v_2$. K is also the static frame; K$'$ and K$''$ move at velocity $v_1$ and $v_2$, respectively.
\end{flushleft}
\end{figure}

From view of the moving frame K$'$, the two electrons are both static, so magnetic field will not be generated, and there will be no Lorentz force. That is to say, there is only electrostatic force between the two electrons.

The electric field that e$_1$ produces at the position of e$_2$ is
\be
E'_y=\frac{1}{4\pi\varepsilon_0}\frac{e}{r^2},~~E'_x=E'_z=0
\ee
Where $e$ is charge of the electron, $r$ is the distance between e$_1$ and e$_2$. The force that e$_2$ feels is
\be
F_0=eE'_y=\frac{1}{4\pi\varepsilon_0}\frac{e^2}{r^2}
\ee
$F_0$ is an electrostatic repulsive force in upward direction.

Because the experimental phenomenon is independent on reference frame, the two electrons being without magnetic force in frame K$'$ will not change their interaction in frame K. On the whole, the interaction between two moving electrons with same velocity is the same as that between two static electrons---there only exists electrostatic force.

Second, consider the case of two electrons with different velocities.

At a certain moment, the relative positions of the two electrons are still the same as given in Fig. 5(a). The velocities of e$_1$ and e$_2$ are $v_1$ and $v_2$, respectively. The velocity difference $v=v_2-v_1$. e$_1$ is still the field charge and e$_2$ still the test charge. K is the static frame. The moving frame K$'$ is fixed to e$_1$, and another moving frame K$''$ is fixed to e$_2$. See Fig. 5(b).

From view of K$'$, the field charge e$_1$ is static and will not generate magnetic field. The test charge e$_2$ will not feel Lorentz force, and there is only electrostatic force between the two electrons. Here the electric field generated by e$_1$ at the position of e$_2$ can still be described with (5.1). The electrostatic force felt by e$_2$ can still be described with (5.2). This force still equals the electrostatic interaction {\bf F}$_0$ between two static electrons.

From the view of frame K$''$, e$_1$ is moving and will simultaneously generate electric field and magnetic field as follows:
\be
E''_y=\frac{E'_y-vB'_z}{\sqrt{1-v^2/c^2}}=\frac{1}{\sqrt{1-v^2/c^2}}\frac{1}{4\pi\varepsilon_0}\frac{e}{r^2}
\ee
\be
B''_z=\frac{B'_z-v/c^2E'_y}{\sqrt{1-v^2/c^2}}
     =-\frac{v/c^2}{\sqrt{1-v^2/c^2}}\frac{1}{4\pi\varepsilon_0}\frac{e}{r^2}
\ee
The other components of {\bf E}$''$ and {\bf B}$''$ are all zero. Because e$_2$ is static relative to K$''$, it will not feel magnetic force. The only force e$_2$ feels is electric force which can be described as follows:
\be
F_1=eE''_y=\frac{1}{\sqrt{1-v^2/c^2}}\frac{1}{4\pi\varepsilon_0}\frac{e^2}{r^2}
\ee

Comparing with (5.2), it can be found that $F_1=F_0/\sqrt{1-v^2/c^2}$. This agrees with the relativistic transformation of a single cross force between two reference frames. Therefore, {\bf F}$_1$ in K$''$ and {\bf F}$_0$ in K$'$ represent the same force, which is still the electrostatic force between two static electrons.

It is derived from the relativity theory that there is only electrostatic force (repulsion) and no Lorentz magnetic attractive force between the electrons. However, many experiments of self-magnetic field \cite{Obenschain: 1979, Yoshikawa: 1971} and self-focusing \cite{Hirano: 2014, Whittum: 1993} of electron beam indicated that there is magnetic force between moving electrons. We call this contradiction as magnetic force paradox.

\section{Summary}

This paper presents four paradoxes about the special theory of relativity: displacement paradox, electromagnetic transformation paradox, Doppler paradox and magnetic paradox. For these paradoxes, the author has not found correct solution. We hope more researchers to pay attention to these problems. If these paradoxes can be correctly solved, they are sure to play a useful role in enriching the theory system of relativity.

\providecommand{\href}[2]{#2}

\end{document}